\documentclass[letterpaper,number,1p]{elsarticle}

\usepackage[T1]{fontenc}
\usepackage[latin9]{inputenc}
\usepackage{amsmath}
\usepackage{amssymb}
\usepackage{esint}
\usepackage{microtype}
\usepackage{subfig}
\usepackage{booktabs}
\usepackage{amsfonts}
\usepackage[mathlines,displaymath]{lineno}
\newcommand{\Bsum}{\sum_{b\in\partial\mathcal{B}}}
\newcommand{\Csum}{\sum_{c\in\partial\mathcal{B}\cup \mathcal{B}}}
\begin{document}

\begin{frontmatter} 
\title{\Large Measurement of higher-order stress-strain
              effects in granular materials
              undergoing non-uniform deformation}
\author[up]{Matthew~R.~Kuhn\corref{cor1}}
  \ead{kuhn@up.edu}
\author[mas]{Ching S. Chang}
  \ead{chang@ecs.umass.edu}
\cortext[cor1]{Corresponding to:
               Donald P. Shiley School of Engineering,
               University of Portland,
               5000 N. Willamette Blvd.,
               Portland, OR, 97203, USA.
               Email: \texttt{kuhn@up.edu}}
\address[up]{Br. Godfrey Vassallo Prof. of Engrg.,
             Donald P. Shiley School of Engrg., Univ. of Portland,
             5000 N. Willamette Blvd., Portland, OR 97231, USA}
\address[mas]{Dept.\ of Civil and Env.\ Engrg., University
              of Massachusetts, Amherst, MA 01002, USA}
\begin{abstract}
Discrete element (DEM) simulations demonstrate that
granular materials are non-simple, meaning that
the incremental stiffness of a granular assembly
depends on the gradients of the strain increment as well as on the
strain increment itself. 
In quasi-static simulations, two-dimensional granular assemblies
were stiffer when the imposed deformation was
non-uniform than for uniform deformation.
The contacts between particles were modeled as
linear--frictional contacts with no contact moments.
The results are interpreted in the context of a higher-order
micro-polar continuum, which admits the possibility
of higher-order stress and couple-stress.
Although the behavior was non-simple,
no evidence was found for a couple-stress or
an associated stiffness.
The experimental results apply consistently to three particle shapes
(circles, ovals, and a non-convex cluster shape),
to assemblies of three sizes (ranging from 250 to 4000 particles),
and at pre-peak and post-peak strains.
\end{abstract}
\end{frontmatter}
\section{\large Introduction}
Large solid regions are commonly modeled as homogeneous continua, ignoring
the underlying heterogeneity that may exist at a smaller scale and, in
the case of granular materials, also
ignoring the discontinuous nature of the micro-scale movements.
Any continuum model of a large macro-region
requires the choice of a continuum
class and of a constitutive form.
The most common continuum class is a
classical, Cauchy continuum, in which the internal force and deformation quantities
are the conventional stress and strain.
Micro-morphic continua,
such as a Cosserat continua, are examples of non-classical continua in which the
conventional strain at a material point is augmented with other kinematic quantities
(micro-rotation, micro-stretch, higher-order micro-strains, etc.), and the
conventional stress is augmented by corresponding, conjugate quantities
(couple-stress, higher-order stress, etc.)
\cite{Mindlin:1964a,Pabst:2005a},
see e.g. \cite{dellIsola:2015a} for a perspective on the origin
of such non-local theories.
The most common constitutive form is that of a simple material, in which
stress at a continuum point depends upon the history of the local strain and its
rate at the point.
With non-simple materials, such as second-gradient materials \cite{Mindlin:1965a},
the stress at a material point depends on the local gradients of
strain.
Non-simple materials also include non-local media in which stress at a
point depends on the deformation within a small region around the point.
\par
With granular materials, the issue of a proper continuum treatment at the
macro-scale is of particular importance, because heterogeneity is inherent and
pervasive and can occur at a scale that is significant when compared with
a specimen's size \cite{Kuhn:2003d,MuirWood:2012a,Wolf:2005a,Misra:2016a}.
Moreover, the strength of a granular material is not only affected by
heterogeneity,
but strength is largely a
bulk expression of material behavior within localized
deformation features such as shear bands, compression bands, and micro-bands
\cite{Desrues:1996a,Kuhn:2010a,Yang:2012a,LeBouil:2014a,Misra:2015a}.
As a consequence, the observed mechanical
behavior of a specimen,
particularly during post-peak softening, is likely affected
by the size of the specimen relative to that of its localization features.
The view of granular materials as simple materials has already been called
into question by experiments that demonstrate an effect of the
gradients of shearing strain on the shear stress \cite{Kuhn:2002a}.
\par
The paper presents evidence of non-simple and size-de\-pen\-dent behavior
during compression and extension loading,
and it is organized in the following manner.
We begin by describing DEM experiments, in which small
rectangular assemblies are deformed
either uniformly or in a non-uniform manner.
The breadth of these assemblies is similar to the observed thickness
of shear bands.
We then present a continuum framework for interpreting the
stress-response to the imposed deformations~---a generalized
micro-morphic continuum~--- and a consistent set of stress
measures that apply to discrete, granular media.
We then analyze the simulation results, determining whether
assembly size affects the stress-response,
whether the results support a Cosserat approach with couple-stress,
and whether the stress--strain response depends upon
the gradients of strain.
\section{Experiments}\label{sec:experiments}
\begin{figure}
  \centering
  \includegraphics[scale=0.97]{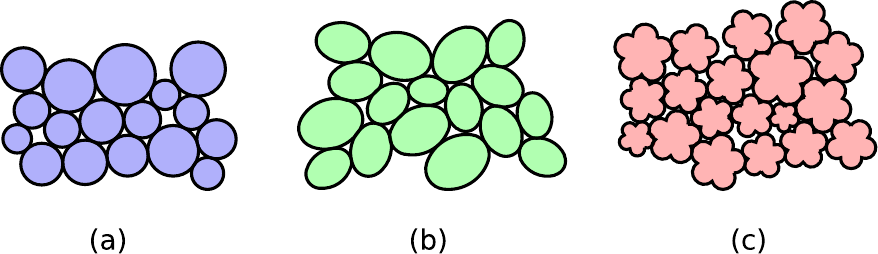} 
  \caption{Three particle shapes used in the simulations:
            circles, ovals, and nobbies.
            \label{fig:shapes}}
\end{figure}

Two series of slow, quasi-static
discrete element (DEM) simulations were conducted on square
two-di\-men\-sional assemblies of particles of three different shapes:
circles, ellipse-like ovals, and composite 
non-convex ``nobby'' shapes that were
formed from five satellite circles arranged around a central circle
(Fig.~\ref{fig:shapes}).
Simulations with the non-circular particles were intended to 
examine possible micro-polar effects that have been conjectured
to result from elongated shapes or from particle pairs that share
multiple contacts \cite{Froiio:2006a}.
Briefly, the two series of simulations
were incremental loadings of the two types
shown in Fig.~\ref{fig:twoseries}:
\begin{figure*}
  \centering
  \includegraphics{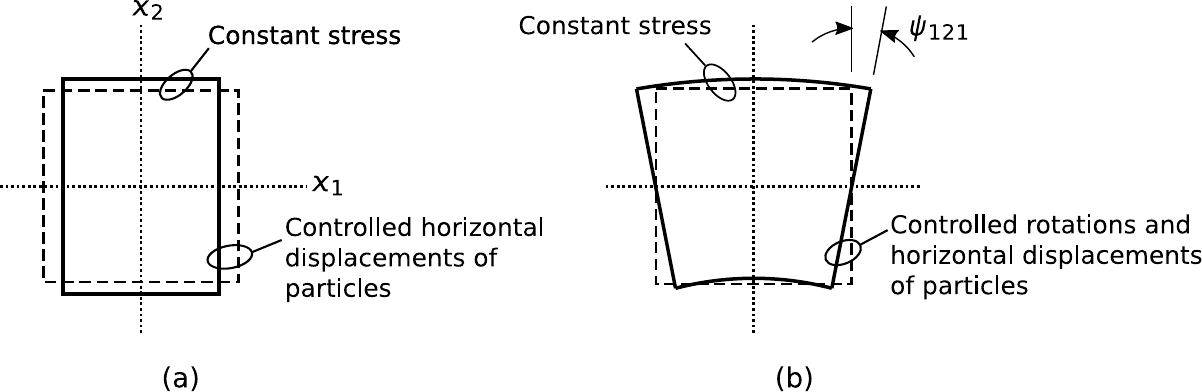}
  \caption{Two series of incremental loadings of 
           particle assemblies:
           (a)~uniform biaxial compression and extension;
           (b)~non-uniform deformation with tilting
           side boundaries.
           \label{fig:twoseries}}
\end{figure*}
(a)~increments of uniform horizontal biaxial compression/extension, and
(b)~increments in which 
a non-uniform bending-type deformation was intentionally imposed.
By comparing the two series of experiments, we determined the
incremental effects
of higher-order gradients of the displacement field
(i.e. gradients of strain) and of the rotation field.
These two series of incremental simulations were conducted after
an initial stage of nearly uniform biaxial compression that
brought the assemblies to four
different initial strain levels, including zero strain,
two strains at and beyond
the condition of peak stress,
and strain at the post-peak critical state.
\par
As a final variable in the simulations,
we conducted simulations on assemblies of different
sizes, ranging from assemblies of only 256 particles to assemblies
of over 4,000 particles.
The stress-strain behavior of such small assemblies,
even those with thousands of particles, can be quite
erratic and can be sensitive
to the initial particle arrangement \cite{Kuhn:2009b},
so we conducted simulations on multiple (as many as 300) initial
assemblies of a given size and then averaged the results.
To summarize, variations in the simulations allowed us
to determine the effects of non-uniform strain (and rotation)
on the incremental stiffnesses of two-dimensional assemblies
and the effects of the three following variables:
particle shape, assembly size, and initial loading strain.
\par
\begin{figure}
  \centering
  \includegraphics{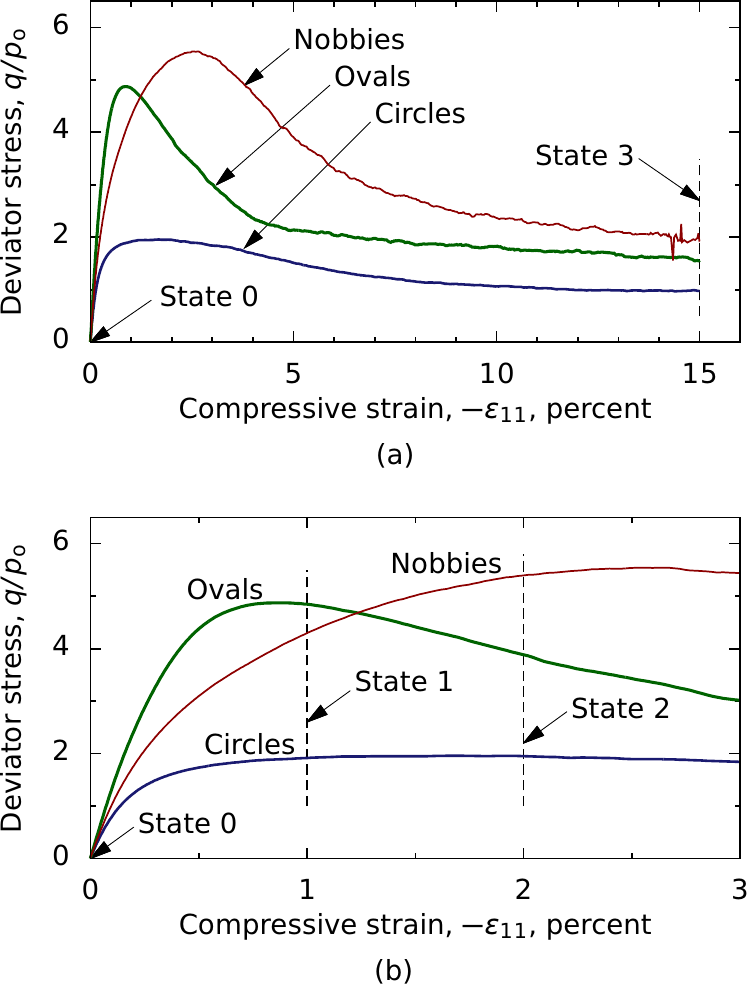}
  \caption{Stress and strain for the initial phase of
           monotonic biaxial compression with constant
           lateral stress.
           The incremental response to a
           non-uniform deformation field was measured
           at four states of strain.\label{fig:Stress}}
\end{figure}

All simulations began with dense isotropic random arrangements
of particles that were contained within periodic boundaries.
The particle sizes, regardless of shape, were poly-disperse,
with a size range of 0.56$D$ to 1.7$D$, where $D$ is the
mean size.
The oval particles had a length/width aspect ratio of
1.30, and the nobby particles were distinctly non-convex,
so that two neighboring particles could touch at one,
two, or three points.
Linear-frictional contacts were used, with equal normal
and tangential stiffnesses $k$ and a friction coefficient
$\mu=0.50$.
No contact moments were applied in the simulations.
Both series of the incremental tests
in Fig.~\ref{fig:twoseries}
followed an initial stage of 
biaxial compression, in which the horizontal width of
an assembly was reduced at a constant rate while maintaining
a constant vertical stress $\sigma_{22}$.
Periodic boundaries were used throughout this initial
loading stage.
Figure~\ref{fig:Stress} shows the results of the initial loading
and the four strains at which the two series of incremental
simulations were conducted.
%
\par
The series of non-uniform ``bending'' simulations 
in Fig.~\ref{fig:twoseries}b could not
be conducted with periodic boundaries, so
after the initial stage of biaxial compression,
an assembly's periodic boundaries were removed and replaced with
an irregular ``flexible'' boundary that passed from center to
center of the peripheral particles around an assembly's
perimeter.
This process, an alternative to using flat platens,
creates assemblies in which the particles
retain their initial arrangement, with a fabric
and stress that is nearly the same as during the
preceding loading stage.
Both series of incremental simulations
in Fig.~\ref{fig:twoseries}
were then conducted with flexible boundaries.
\par
After the initial stage of loading to a
particular strain state,
the first series of simulations,
shown in  Fig.~\ref{fig:twoseries}a, were
of conventional biaxial compression and extension:
each assembly was horizontally compressed (or extended)
between its two
sides of boundary particles, which approached (or retracted)
at a constant rate of strain, while maintaining a
constant stress along
the upper and lower sides.
These simulations
were performed to determine
the incremental Young's modulus $E$
for conditions of loading and unloading,
and they provide a benchmark against which the second series of
tests can be compared.
In a sense, the incremental simulations were a continuation
of the initial monotonic loading, although with flexible
rather than periodic boundaries.
As will be seen, the non-uniform ``bending'' experiments
of Fig.~\ref{fig:twoseries}b induced
unloading within a part of an assembly, so we also
conducted the incremental \emph{extension}
simulations to determine the
unloading modulus.
%
%
\par
The second series of experiments were special
bending tests, in which the left and right sides
of and assembly
(i.e., chains of particles)
were rotated as shown in Fig.~\ref{fig:twoseries}b.
To induce these conditions, particles along the left
and right sides were displaced horizontally and were forced
to corotate with their boundary.
These tests provide a means of measuring the 
material response to imposed gradients of strain 
and of particle rotation.
As with the compression tests, a constant stress
was maintained along the top and bottom boundaries during the bending
tests.
\par
In both the compression and bending series
of simulations,
the top and bottom boundaries
were flexible, with a uniform vertical pressure applied
to virtual links that joined the centers of neighboring
peripheral boundary
particles, in the manner of a flexible membrane in geotechnical
testing.
The peripheral particles along these boundaries could
freely move and rotate under the influence of a constant
boundary stress $\sigma_{22}$ applied at the particles' centers
(see \cite{Bardet:1991a} for the first known use of such boundaries).
Along the left and right boundaries in both sets of experiments,
the vertical movements of particles were unconstrained, but
the horizontal movement $u_{1}^{p}$
and rotation $\theta_{3}^{p}$ of each ``$p$'' particle
was constrained to conform with 
an average strain $\varepsilon_{11}$ and
an average strain gradient $\psi_{112}$:
\begin{align}\label{eq:displacement}
  du_{1}^{p} &= d\varepsilon_{11}^{p} x_{1}^{p}
               +d\psi_{112} x_{1}^{p} x_{2}^{p}\\ \label{eq:rotation}
  d\theta_{3}^{p} &= -d\psi_{112} x_{1}^{p}
\end{align}
where $d\varepsilon_{11}$ is the conventional
horizontal strain increment,
$d\psi_{112}$ is the vertical gradient of the horizontal strain,
and $x_{i}^{p}$ is the location of boundary particle $p$'s center.
In Eq.~(\ref{eq:displacement}),
the deformation $d\psi_{112}$ is the second-gradient of displacement,
$du_{1,12}$, applied at the boundaries.
We emphasize that
Eqs.~(\ref{eq:displacement})--(\ref{eq:rotation})
are the imposed conditions along the side boundaries,
but the two deformations, $d\varepsilon_{11}$ and $d\psi_{112}$,
can also be considered ``macro-strains'':
characteristic strains within a non-homogeneous
micro-region that constitutes a representative volume element
(RVE, Section~\ref{sec:continuum}).
To constrain the movements and rotations of the side
particles, restraining forces and moments were applied
to the centers of these boundary particles.
The combined effect of these external forces and moments can be
computed as equivalent ``bending moments'' applied to the
left and right sides, as will be discussed later.
Unlike the side particles, the assembly's interior
particles were free to both move and rotate, as they
accommodated the imposed boundary conditions.
\par
\begin{figure*}
  \centering
  \includegraphics{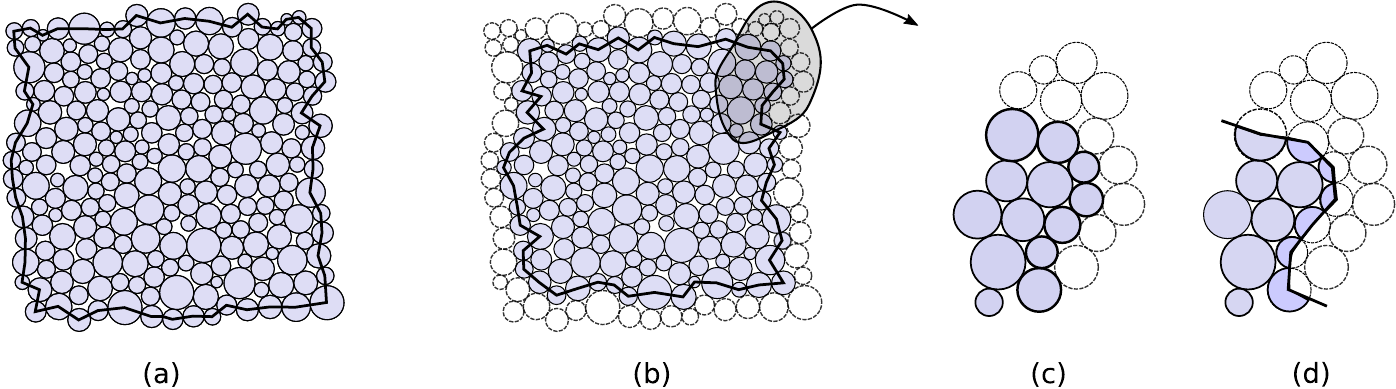}
  \caption{Alternative RVEs within an assembly of
           256 disks: (a)~assembly of all disks,
           noting that the movements of
           side particles were imposed;
           and (b)~sub\--assembly for computing
           the response a freely responding RVE.
           These sub-assemblies were used to measure
           the stress-response in the paper.
           For the sub-assembly, two RVE boundaries
           are possible:
           (c)~boundary fully encompassing the peripheral
           particles, and
           (d)~boundary passing through the centers of
           peripheral particles.
           \label{fig:subassembly}}
\end{figure*}

With each simulations
in Fig.~\ref{fig:twoseries},
our intent was to determine the stress response
of a granular RVE that was freely responding to
a small, incremental deformation.
Because rotations and movements were imposed
upon those particles along the sides of an assembly
and constant force was applied to those particles along the
top and bottom,
these boundary particles are not considered part of a
``freely responding'' RVE (Fig.~\ref{fig:subassembly}a).
For this reason, we measured the response
of an interior sub-assembly that was fully
contained within (and surrounded by) the boundary particles
(Fig.~\ref{fig:subassembly}b).
The manner in which stress was computed within this
freely responding sub-assembly is described in the next
section, which also describes two variations of analyzing
the sub-assemblies depicted in Fig.~\ref{fig:subassembly}b.
Note that stress and deformation are computed within sub-assemblies
that contain fewer particles than the full assembly:
the assemblies of 256, 1024, and 4096 particles
encompassed sub-assemblies of about 180, 900, and 3800
particles, respectively.
%
%
%
%
\section{Generalized continuum setting}\label{sec:continuum}
We computed various stress quantities within the assemblies
by taking the theoretic approach
advocated by Germain \cite{Germain:1973a},
in which movements and strains are accepted as the privileged,
fundamental quantities,
and stresses, in their various forms, are
merely derived as conjugates to these movements and strains.
If the operative displacement quantities
\emph{at a continuum point}
are those of a generalized, micro-polar continuum
and
are taken as the displacement gradient
$\delta u_{i,j}$, second gradient
$\delta u_{i,jk}$, micro-rotation $\delta \theta_{i}$,
and micro-rotation gradient $\delta\theta_{i,j}$,
then the internal virtual work $\delta W^{\text{a}}$
that is formed from these quantities
is (see \cite{Chang:2005a})
\begin{equation}
\label{eq:WIa}
\delta W^{\text{a}} = \sigma_{ji}\delta u_{i,j}
                    + \sigma_{jki}\delta u_{i,jk}
                    + T_{i}\delta\theta_{i}
                    + T_{ji}\delta\theta_{i,j}
\end{equation}
In this expression of virtual work, force quantity
$\sigma_{ji}$ is immediately recognized as the stress,
as it is conjugate with the displacement gradient,
although the equation imposes no condition of symmetry on this stress.
The other force quantities are the higher-order
stress $\sigma_{jki}$ and the internal torque
and internal torque-stress,
$T_{i}$ and $T_{ij}$.
Although no continuum internal torques $T_{i}$ apply in
our simulations, the higher-order stress
and torque stress were measured in a manner that is described
later.
The internal virtual work of Eq.~(\ref{eq:WIa})
can be rearranged to yield 
the more familiar form of a higher-order Cosserat continuum:
\begin{equation}
\label{eq:WIb}
\delta W^{\text{b}} = \sigma_{ji}(\delta u_{i,j} + e_{ijk}\delta\theta_{k})
     + \sigma_{jki}(\delta u_{i,jk} + e_{ij\ell}\delta\theta_{\ell,k})
     + \mu_{ji} \delta\theta_{i,j}
\end{equation}
where we have introduced the
couple-stress $\mu_{ji}$, which is complementary
with the gradient of the micro rotation
$\delta\theta_{i,j}$.
Stress $\sigma_{ji}$ can be
asymmetric, and its non-symmetric part is complementary
with the difference between micro-rotation
$\delta\theta_{k}$ and the asymmetric
part of the displacement gradient
$\delta u_{i,j}$.
The virtual works of
Eqs.~(\ref{eq:WIa}) and~(\ref{eq:WIb})
are equivalent,
so that the couple-stress,
torque-stress, and higher-order
stress are related, as
\begin{equation}\label{eq:muTsigma}
\mu_{ji} = T_{ji} - e_{k\ell i}\sigma_{\ell jk} \;.
\end{equation}
%
%
\par
The stress quantities in
Eqs.~(\ref{eq:WIa})--(\ref{eq:muTsigma}) apply to a point within a
continuum. 
When the continuum is intended to represent a discrete, granular
material, we use the superscript ``0'' to designate the
\emph{macro-stress} of a small granular region
(i.e., a \emph{micro-region} or RVE) that
is representative of a continuum point \cite{Chang:2005a}.
That is, macro-stresses are the representative stresses
of a small granular region rather than
of a continuum point or even of a point within an individual
particle inside the region.
\par
The various
macro-stresses are measured
for a micro-region~--- an aggregate of discrete granular constituents
which forms the RVE of a continuum point.
The macro-stresses that arise from the continuum settings of
Eq.~(\ref{eq:WIa}) and~(\ref{eq:WIb})
can be computed within a granular micro-region $\mathcal{B}$
from summations
of contact forces and contact moments, so that the virtual
works of the continuum stresses and of the macro-stresses
coincide (see \cite{Chang:2005a}).
Each macro-stress is computed from the contact forces,
with either of two alternative sums.
One sum involves the contacts $b\in\partial\mathcal{B}$
between peripheral, boundary particles and the RVE's exterior;
the second sum is of the contacts
$c$ among the interior and boundary particles.
The alternative expressions, using either the boundary 
contacts or all contacts, are as follows:
\begin{align}
\label{eq:sigma}
\sigma_{ji}^{0} &= \frac{1}{V} \Bsum f_{i}^{b} x_{j}^{b} =
                   \frac{1}{V} \Csum f_{i}^{c} l_{j}^{c}   \\
\label{eq:sigmaH}
\sigma_{jki}^{0} &= \frac{1}{2V} \Bsum f_{i}^{b} x_{j}^{b} x_{k}^{b} =
                    \frac{1}{2V} \Csum f_{i}^{c} J_{jk}^{c} \\
\label{eq:mu}
\mu_{ji}^{0}      &= \frac{1}{V} \Bsum \left( m_{i}^{b} x_{j}^{b}
                      -\frac{1}{2}e_{ik\ell}f_{k}^{b} x_{\ell}^{b} x_{j}^{b}
                      \right) \\ \notag
                 &=
                    \frac{1}{V} \Csum \left[ m_{i}^{c} l_{j}^{c} + e_{ik\ell}f_{k}^{c}
                      \left(\frac{1}{2}J_{\ell j}^{c} 
                            - x_{\ell}^{c} l_{j}^{c}\right)\right] \\
\label{eq:T}
T_{ji}^{0} &= \frac{1}{V} \Bsum m_{i}^{b} x_{j}^{b} \\ \notag
           &= \frac{1}{V} \Csum \left[
                        m_{i}^{c} l_{j}^{c} + e_{ik\ell}f_{k}^{c}
                        (J_{\ell j}^{c} - x_{\ell}^{c} l_{j}^{c})
                        \right]
\end{align}
In these expressions,
$f_{i}^{b}$ are external boundary forces applied to the peripheral
particles of a
micro-region;
$f_{i}^{c}$ are internal
contact forces between particles within the region
and external contact forces between
peripheral particles and the region's exterior; 
$m_{i}^{b}$ and $m_{i}^{c}$ are boundary moments and contact moments;
$l_{j}$ are branch vectors that join the centers of contacting particle pairs
or join the centers of peripheral particles and exterior contact points; 
$x_{j}^{b}$ and $x_{j}^{c}$ are the locations
of external forces or of internal contacts; 
and tensors $J_{jk}^{c}$ are the quadratic differences 
$x_{j}^{q}x_{k}^{q} - x_{j}^{p}x_{k}^{p}$ for two contacting particles,
$p$ and $q$, or for peripheral particles and their exterior contacts.
In a two-dimensional setting, $V$ is the RVE area;
with three-dimensional RVEs, $V$ is the volume.
Each equation gives two summations for a macro-stress quantity:
the first
summation is of peripheral contacts ``$b$'' along the boundary
$\partial\mathcal{B}$ of region $\mathcal{B}$; whereas, the second
summation is of contacts ``$c$'' both within the region
and between the region and its exterior,
$c\in \partial\mathcal{B} \cup\mathcal{B}$.
Because the DEM algorithm uses a relaxation technique that
only achieves an approximate equilibrium, the external and
internal sums in Eqs.~(\ref{eq:sigma})--(\ref{eq:T}) were not precisely
equal, but they never differed by more than 0.1\%.
\par
As was described in Section~\ref{sec:experiments},
Fig.~\ref{fig:subassembly}b depicts a sub-assem\-bly
RVE that is
surrounded by an external layer of particles whose movements and
rotations (or external forces) were controlled during
a simulated increment of deformation,
as in Eqs.~(\ref{eq:displacement})--(\ref{eq:rotation}).
The expressions of macro-stress in Eqs.~(\ref{eq:sigma})--(\ref{eq:T})
suggest two approaches to computing stress within
a sub-assembly RVE, which are illustrated in
Figs.~\ref{fig:subassembly}c and~\ref{fig:subassembly}d.
Both figures show three sets of particles:
unshaded external particles that are outside the RVE,
peripheral particles of the RVE,
and interior particles fully inside the RVE.
With the variant of Fig.~\ref{fig:subassembly}c,
the RVE fully encompasses the
peripheral particles;
whereas, the boundary of the RVE in
Fig.~\ref{fig:subassembly}d passes through
the peripheral particles but includes the interior portions
of these particles along with the contacts between the peripheral
and other RVE particles.
\par
Eqs.~(\ref{eq:sigma})--(\ref{eq:T}) apply differently to the
two boundary variants in
Figs.~\ref{fig:subassembly}c and~\ref{fig:subassembly}d.
With Fig.~\ref{fig:subassembly}c, the boundary forces and moments,
$f^{b}_{i}$ and $m^{b}_{i}$, are the contact forces and moments
between the peripheral and external particles.
Because no contact moments were present in our simulations,
the torque-stress $T^{0}_{ji}$ is zero
\cite{Chang:2005a}.
Moreover, the contribution
$e_{ik\ell}f_{k}^{b} x_{\ell}^{b} x_{j}^{b}$
in Eq.~(\ref{eq:mu}) is zero for two-dimensional assemblies
(see Eq.~\ref{eq:Kmu} below), so the couple-stress
$\mu^{0}_{ji}$ in Eq.~(\ref{eq:muTsigma}) is also zero.
The situation is different for the RVE in
Fig.~\ref{fig:subassembly}d.
Boundary forces and moments,
$f^{b}_{i}$ and $m^{b}_{i}$, are applied to the bodies
of peripheral particles,
as these body forces must balance the contact forces
between the peripheral and interior RVE particles.
For this choice of boundary,
the moments $m^{b}_{i}$ are not necessarily zero,
and as a result,
torque-stresses and couple-stresses
can arise with this RVE.
\section{Results}\label{sec:results}
Our simulations were intended to measure
the \emph{incremental} macro-stress response to
small deformation increments: specifically, the increments
$d\varepsilon_{11}=du_{1,1}$
and $d\psi_{112}=du_{1,12}$
(see Eqs.~\ref{eq:displacement} and~\ref{eq:rotation}).
Two factors must be considered when computing the incremental
response of an assembly in the context of the macro-stresses
of Eqs.~(\ref{eq:sigma})--(\ref{eq:T}).
First,
it is clear from the external contact expressions for
$\sigma_{jki}^{0}$ and $\mu_{ji}^{0}$,
which contain the products $x_{j}x_{i}$,
that these two macro-stresses depend upon the size
of the micro-region that is begin considered:
doubling the assembly size will double the contributions of these
products to their macro-stresses
(for example, with the first stress measure in Eq.~\ref{eq:sigmaH},
doubling a 2D assembly's length and width but maintaining
the same particles' sizes will double the number of particles
along the assembly's perimeter, quadruple the area $A$, and quadruple
the $x^{b}_{i}x^{b}_j$ products, thus increasing
$\sigma^{0}_{jki}$ by a factor of two).
Second, the incremental stiffness moduli of an assembly
(micro-region) will depend upon (and will be roughly proportional to)
the contact stiffness, $k$, between particles.
To analyze the results of our two-dimensional simulations
in a consistent,
size-independent manner, we normalized the results with
the following dimensionless stiffness moduli:
\begin{align}\label{eq:Kepsilon}
K_{\varepsilon} &=
   \frac{1}{k} \left(d\sigma^{0}_{11}/d\varepsilon_{11}\right)\\
\label{eq:Kpsi}
K_{\psi} &=
\frac{2}{k} \left(d\sigma^{0}_{121}/d\psi_{121}\right)
\frac{A}{I_{22}}\\
\label{eq:Ktheta}
K_{\theta} &=
\frac{1}{k} \left(dT^{0}_{13}/d\theta_{3,1}\right)
\frac{A}{I_{22}}
= K_{\varepsilon}\frac{A}{I_{22}} \ell^{2}
\end{align}
%
where we explicitly represent the volume $V$
of a two-dimensional assembly
as it's area $A$.
These three moduli are the stiffnesses associated with the 
first, second, and fourth terms on the right of
Eq.~(\ref{eq:WIa}).
(In a three-dimensional setting, each modulus would
be divided by a micro-scale measure with dimensions of length,
for example, the mean particle size $D$,
to maintain a dimensionless character.)
Note that the experiments are limited to determining three of
the many moduli that are associated with
stresses $\sigma^{0}_{ij}$, $\sigma^{0}_{ijk}$,
and $T^{0}_{ij}$, and further experiments would be required
for examining the other moduli
(see \cite{Placidi:2015a,Placidi:2017a}). 
\par
The first modulus, $K_{\varepsilon}$,
is simply the normalized Young's modulus
that we measured with the series of incremental biaxial compression
and extension tests with constant lateral stress
(see Fig.~\ref{fig:twoseries}a).
\par
\begin{figure}
  \centering
  \includegraphics{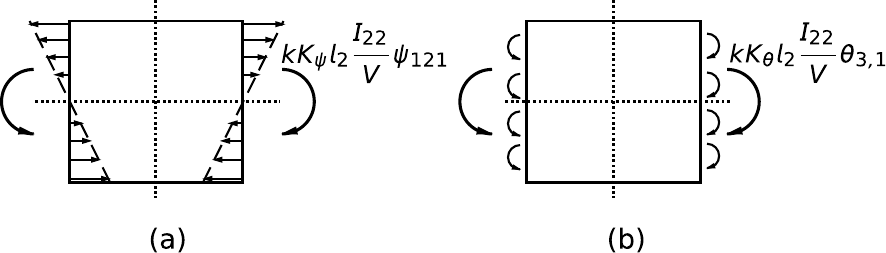}
  \caption{Stiffnesses attributed to boundary forces
           and moments: (a)~the strain gradient
           $d\psi_{121}=du_{1,21}$ produces non-uniform
           side forces with associated stiffness
           $K_{\psi}$;
           (b)~the rotation gradient $d\theta_{3,1}$
           produces
           side moments with associated stiffness
           $K_{\theta}$.
           Parameter $l_{2}$ is the assembly height.
           \label{fig:twoKs}}
\end{figure}

The modulus $K_{\psi}$ is the Young's modulus of a
two-di\-men\-sion\-al linear-elastic rectangular solid
undergoing non-uniform deformation,
as derived by treating the rectangle
as an Euler--Bernoulli beam, where $d\psi_{121}$
is the beam's curvature (Fig.~\ref{fig:twoKs}a),
and $I_{22}=\int x_{2}x_{2}\,dA$
is the second-moment of the assembly's volume
measured about the $x_{1}$ axis.
If the material is simple, such that stress is independent
of the gradients of strain, then the two moduli,
$K_{\varepsilon}$ and $K_{\psi}$, will be equal.
\par
The third modulus $K_{\theta}$ is the stiffness
associated with the applied boundary torques that
are required to rotate particles along the side boundaries
(Fig.~\ref{fig:twoKs}b).
This modulus is that of an elastic two-dimensional Cosserat plate,
in which rotational stiffness is exclusively derived from the
couple-stress \cite{Vardoulakis:2009b}. 
This rotation stiffness is commonly expressed as the product
of the Young's modulus and a
squared measure of micro-mechanical length $\ell$.
\par
A fourth stiffness is associated
with the couple-stress $\mu_{13}$, noting the
relationship in Eq.~(\ref{eq:muTsigma}),
\begin{align}\label{eq:Kmu}
K_{\mu} &=
\frac{1}{k} \left(d\mu^{0}_{13}/d\theta_{3,1}\right)
\frac{A}{I_{22}}\\
&=K_{\theta} + \frac{1}{k}
 \left(
 d\sigma^{0}_{312}/d\theta_{3,1} -
 d\sigma^{0}_{213}/d\theta_{3,1}
 \right)\frac{A}{I_{22}}
\end{align}
In our two-dimensional setting, the stresses
$\sigma^{0}_{312}$ and $\sigma^{0}_{213}$ are zero,
so that moduli $K_{\mu}$ and $K_{\theta}$ are equivalent.
Only the latter is reported below.
\par
The stiffnesses in Eqs.~(\ref{eq:Ktheta}) and~(\ref{eq:Kmu})
can be measured with either of the two types of RVE boundaries
illustrated in Figs.~\ref{fig:subassembly}c and~\ref{fig:subassembly}d,
by computing the small changes in the various stresses
that resulted from the displacement probes of
the two series of simulations
(see Figs.~\ref{fig:twoseries}a and~\ref{fig:twoseries}b).
We found that the two types of boundaries gave nearly identical results,
and only the results with the boundaries shown in
Fig.~\ref{fig:subassembly}d are reported herein.
\subsection{Response to biaxial compression and extension}
Recall that our incremental simulations followed periods
of sustained, monotonic horizontal biaxial compression.
Our primary interest is the response of assemblies
to the non-uniform ``bending'' deformation of Fig.~\ref{fig:twoseries}b,
but we also conducted simulations of increments of
biaxial compression and extension (Fig.~\ref{fig:twoseries}a),
as these simulations will serve as the reference condition
with which the response to non-uniform deformation is compared.
The non-uniform simulations of Fig.~\ref{fig:twoseries}b
produced horizontal
compressive loading in the lower half of the assembly while producing
a reversed, extensional increment in the upper half
(Fig.~\ref{fig:twoseries}b),
calling for both biaxial compression and biaxial extension simulations
to serve as dual reference conditions.
The normalized compression and extension moduli are designated
as $K_{\varepsilon}^{\text{load}}$
and $K_{\varepsilon}^{\text{unload}}$.
\par
Table~\ref{table:Ksigma1} presents $K_{\varepsilon}$ values
for the three particle shapes at four different strains
for the assemblies containing 1024 particles.
\begin{table}
  \centering
  \caption{Moduli $K_{\varepsilon}$ for incremental
           loading and unloading of assemblies of 1024
           particles.
           \label{table:Ksigma1}}
  \begin{tabular}{lccc}
  \toprule
  Shape & Strain & $K_{\varepsilon}^{\text{load}}$
        & $K_{\varepsilon}^{\text{unload}}$\\
  \midrule
  Circles & 0\%   & 0.855 & 0.855\\
          & 1\% & 0.181 & 0.425\\
          & 2\% & 0.173 & 0.335\\
          & 15\% & 0.003 & 0.212\\
  \midrule
  Ovals & 0\% & 1.550 & 1.550\\
        & 1\% & 0.823 & 1.067\\
        & 2\% & 0.715 & 0.695\\
        & 15\% & 0.057 & 0.438\\
  \midrule
  Nobbies & 0\% & 1.095 & 1.095\\
          & 1\% & 0.119 & 0.958\\
          & 2\% & 0.052 & 0.867\\
          & 15\% & 0.046 & 0.476\\
  \bottomrule
  \end{tabular}
\end{table} 
As would be expected, the normalized Young's modulus $K_{\varepsilon}$,
measured with increments of biaxial compression,
decreases with increasing strain, as the material softens
with continued loading.
The loading modulus was smallest for disks and largest
for the assemblies with oval particles.
The fact that the stiffness is greater for ovals than for
the non-convex nobbies indicates that particle elongation
(non-sphericity) has a greater stiffening effect than
particle non-convexity (a type of angularity).
The figure also gives the unloading moduli that were measured
with increments of biaxial extension.
As would be expected during strain hardening,
the unloading modulus
is typically larger than the loading modulus.
The difference is most pronounced for the nobbies and is
least apparent with ovals.
The degradation of the unloading modulus is largely
due to a loss of inter-particle contacts,
as our assemblies dilated during the initial phase
of monotonic loading.
\par
The effect of assembly size is shown in
Table~\ref{table:Ksigma2}, which gives the moduli
$K_{\varepsilon}$ for three assembly
sizes at a common strain of 1\%.
\begin{table}
  \centering
  \caption{Effect of assembly size on modulus $K_{\varepsilon}$
           for assemblies at 1\% strain.
           \label{table:Ksigma2}}
  \begin{tabular}{lccc}
     \toprule
     Shape & Size & $K_{\varepsilon}^{\text{load}}$
           & $K_{\varepsilon}^{\text{unload}}$\\
     \midrule
     Circles &  256 & 0.280 & 0.542\\
             & 1024 & 0.181 & 0.425\\
             & 4096 & 0.107 & 0.396\\
     \midrule
     Ovals   &  256 & 0.953 & 1.072\\
             & 1024 & 0.823 & 1.067\\
             & 4096 & 0.585 & 1.102\\
     \midrule
     Nobbies  & 256 & 0.117 & 1.106\\
             & 1024 & 0.119 & 0.958\\
             & 4096 & 0.110 & 0.884\\
     \bottomrule
  \end{tabular}
\end{table}
Stiffness, whether for loading or unloading,
generally decreases with assembly size, a result observed
at three strains: 1\%, 2\%, and 15\%.
Apart from an inherent size-effect,
this result could be due, in part, to
differences in the fabrics of the differently sized
assemblies.
The process of creating the initial assemblies~--- fitting
them into a small periodic box~---
produces different
densities of contacts, depending upon the numbers of
particles in the original assembly.
Because each additional contact brings an increase in
assembly stiffness,
we should adjust the stiffnesses
in Table~\ref{table:Ksigma2} by
dividing by the corresponding contact densities
(numbers of contacts divided by assembly volume).
With this adjustment, the unloading modulus
$K_{\varepsilon}^{\text{unload}}$ was nearly constant
for the three assembly sizes,
but the loading modulus
$K_{\varepsilon}^{\text{load}}$ still exhibited
the same trend of a decrease in stiffness with
increasing assembly size.
%
\subsection{Response to non-uniform deformation and rotation}
With the simulations
of non-uniform deformation (Fig.~\ref{fig:twoseries}b),
we sought answers to two questions.
First, we determined
whether a stiffness $K_{\theta}$ is associated
with torque-stress (or couple-stress)
for a granular material subject to a field of non-uniform
micro-rotation, $\theta_{3,1}$.
The presence of such
stiffness is central to modeling granular materials
as Cosserat continua.
Second, we determined whether stiffness is greater for
conditions of non-uniform deformation,
with $du_{1,12}\neq 0$, than
for uniform deformation.
We note that the possibility of a non-symmetric
stress, with $\sigma_{12}\neq \sigma_{21}$,
could not be resolved with these simulations,
due to symmetry
of the loading about the $x_{1}$
(Fig.~\ref{fig:twoseries}b).
\par
In regard to the first question,
by applying the displacements and rotations
of Eqs.~(\ref{eq:displacement}) and~(\ref{eq:rotation}),
we produced increments of non-uniform strain and rotation
within the assemblies.
These constraints on displacement and rotation were
applied to the side boundaries
(the unshaded side particles in Fig.~\ref{fig:subassembly}b),
while the stress and the
average deformation and rotation fields were computed for
the interior particles (the shaded RVE particles in
Figs.~\ref{fig:subassembly}b--\ref{fig:subassembly}d).
We found that the average rotation gradient of the
interior particles, $d\theta_{3,1}$, was consistent
with the imposed deformation gradient $d\psi_{112}$
that was applied
at the boundary, and the measured internal
gradient $d\theta_{3,1}$
was typically within 10\% of $d\psi_{112}$.
As was explained in the previous section,
a subtle difference in the side boundaries will either
admit or abrogate moments $m_{3}$ among the peripheral
particles.
If the RVE of Fig.~\ref{fig:subassembly}c is used,
the moments are zero, the torque-stress $T_{13}$ is zero,
modulus $K_{\theta}$ is zero, and the micro-mechanical length
scale $\ell$ is zero.
On the other hand,
boundary moments can exist and can be measured with
the RVE illustrated in Fig.~\ref{fig:subassembly}d,
for which body moments arise to provide moment
equilibrium for the peripheral particles.
We found, however, that
even in the presence of a rotation gradient $d\theta_{3,1}$,
the average torque-stress $T_{13}$ was nearly zero
for the RVE with body moments.
That is,
although these moments were present along an assembly's
peripheral particles, they nearly canceled each other,
leaving a net torque-stress $T_{13}$ close to zero.
This observation was determined
for all particle shapes, all assembly sizes, and at all strains:
the stiffness $K_{\theta}$ in
Eq.~(\ref{eq:Ktheta}) was typically less than 0.001.
These small values correspond to a micro-mechanical length $\ell$
that is much smaller than the mean
particle size~--- a length scale smaller
than the grains themselves.
Because this result applies to all three particle shapes,
we conclude that a micro-rotation stiffness is not induced
by multiple inter-particle contacts (as with the nobby shapes)
or by an elongated particle shape (as with ovals).
\par
To address the second question~--- whether stress is
independent of strain gradients~---
we measured the normalized
modulus $K_{\psi}$ in Eq.~(\ref{eq:Kpsi}).
This modulus is the Young's modulus
that would be consistent with a simple linear-elastic
plate, in which stiffness is \emph{independent}
of any gradients of strain.
If the response of an assembly depends only on
the strain $\varepsilon_{11}$ and is independent
of the strain gradient $d\psi_{112}$,
the two moduli, $K_{\varepsilon}$ and $K_{\psi}$,
will be equal.
We note, however, that
particles in the upper and lower halves
of an assembly are subject to increments
of horizontal tension and compression, respectively
(see Fig.~\ref{fig:twoKs}a), and
we must account, of course, for the different moduli~---
$K_{\varepsilon}^{\text{load}}$ and $K_{\varepsilon}^{\text{unload}}$~---
that apply to the lower and upper parts of an assembly.
That is, we should compare modulus
$K_{\psi}$ with the \emph{average} of the
two incremental stiffnesses,
$K_{\varepsilon}^{\text{load}}$ and $K_{\varepsilon}^{\text{unload}}$,
since the two stiffnesses apply to equal volumes of material.
\par
These comparisons are made in Table~\ref{table:Kpsi1} for
assemblies of three particle shapes and at four strains.
\begin{table}
  \centering
  \caption{Comparison of stiffness,
           $K_{\varepsilon}$ and $K_{\psi}$,
           for biaxial and bending types of deformation,
           for assemblies of 1024 particles.
           \label{table:Kpsi1}}
  \begin{tabular}{lccc}
     \toprule
     Shape & Strain & $\frac{1}{2}
                     (K_{\varepsilon}^{\text{load}}
                     +K_{\varepsilon}^{\text{unload}})$
           & $K_{\psi}$\\
     \midrule
  Circles & 0\%  & 0.855 & 0.806\\
          & 1\%  & 0.303 & 0.547\\
          & 2\%  & 0.254 & 0.398\\
          & 15\% & 0.107 & 0.284\\
  \midrule
  Ovals & 0\% & 1.550 & 1.554\\
        & 1\% & 0.945 & 1.215\\
        & 2\% & 0.704 & 0.756\\
        & 15\% & 0.248 & 0.427\\
  \midrule
  Nobbies & 0\% & 1.095 & 1.035\\
          & 1\% & 0.538 & 0.720\\
          & 2\% & 0.459 & 0.777\\
          & 15\% & 0.261 & 0.367\\
     \bottomrule
   \end{tabular}
\end{table}
Except at the start of loading (state~0 in Fig.~\ref{fig:Stress}),
the modulus $K_{\psi}$ is consistently larger than the
averaged modulus $K_{\varepsilon}$:
the Young's modulus of a granular material is larger
for conditions of non-uniform deformation than for uniform
biaxial loading.
This experimental result demonstrates that granular
materials are non-simple~--- that the stress
depends upon the strain gradients and not solely
on the strain~--- and that a non-simple constitutive
form should be adopted for granular materials.
%
\par
Table~\ref{table:Kpsi2} gives the higher-order modulus
$K_{\psi}$ for assemblies of three sizes:  256, 1024, and 4096
particles.
The results are the incremental stiffnesses for assemblies
that were pre-loaded to a strain of 1\%
(state~1, Fig.~\ref{fig:Stress}).
\begin{table}
  \centering
  \caption{Effect of assembly size on moduli $K_{\varepsilon}$
           and $K_{\psi}$
           for assemblies at 1\% strain.
           \label{table:Kpsi2}}
  \begin{tabular}{lccc}
     \toprule
     Shape & Size & $\frac{1}{2}
                     (K_{\varepsilon}^{\text{load}}
                     +K_{\varepsilon}^{\text{unload}})$
           & $K_{\psi}$\\
     \midrule
     Circles &  256 & 0.411 & 0.675\\
             & 1024 & 0.303 & 0.547\\
             & 4096 & 0.252 & 0.444\\
     \midrule
     Ovals   &  256 & 1.012 & 1.334\\
             & 1024 & 0.945 & 1.215\\
             & 4096 & 0.843 & 1.087\\
     \midrule
     Nobbies  & 256 & 0.612 & 0.832\\
             & 1024 & 0.538 & 0.720\\
             & 4096 & 0.497 & 0.673\\
     \bottomrule
  \end{tabular}
\end{table}
For all three particle shapes and all assembly sizes,
the modulus $K_{\psi}$ is larger than the averaged
Young's modulus $K_{\varepsilon}$.
The ratio of the two moduli, $K_{\psi}/K_{\varepsilon}$,
is largest for the assemblies of disks and smallest for
the oval assemblies.
Moreover, for each particle shape, the ratio
is about the same for all three assembly sizes.
One might think that the effect of assembly
size should diminish with increasing size,
but recall that with each assembly size,
we imposed boundary conditions that produced a uniform
\emph{gradient} of strain across an assembly's full height~---
regardless of the number of particles in the height~---
and the results show that stiffness is augmented by
the presence of this strain gradient,
independent of the assembly size.
\section{Conclusions}
The experiments reveal evidence of an internal length-scale
for granular materials.
This result is manifest in the effect of assembly size on
the loading stiffness:  larger assemblies exhibit softer behavior.
Although this size effect is absent at the start of
loading and in the unloading modulus
(after adjusting for contact density),
the stiffer loading behavior of small samples suggests
that some form of length-scale is operative.
But how should one model this effect?
The measured absence of a Cosserat stiffness~---
a stiffness associated with the spatial gradient of
micro-scale rotation~---
should discourage the use of micro-polar continuum
models for granular materials.
On the other hand, we found that granular assemblies
are much stiffer when subject to a non-uniform deformation
field than would be predicted with a
simple size-independent Young's modulus.
Moreover, this increase in stiffness 
for non-uniform loading conditions was nearly independent
of the assembly size, provided that the imposed second gradient
of displacement (i.e., first gradient of strain) was the
same for each sample size.
These conclusions have only been tested for two-dimensional
granular materials, for a simple linear-frictional contact type,
and for three particle shapes,
but the results clearly
suggests that a length-scale should be directly
incorporated within the constitutive form:
that granular materials should be modeled as non-simple
materials,
in which stress and stiffness depend upon both strain
and the gradients of strain.
%
%
\par
\pagebreak
\bibliographystyle{elsart-num-sort}

\end{document}